\documentclass[conference]{IEEEtran}
\IEEEoverridecommandlockouts

\usepackage{cite}
\usepackage{amsmath,amssymb,amsfonts}
\usepackage{algorithmic}
\usepackage{graphicx}
\usepackage{textcomp}
\usepackage{booktabs}
\usepackage{xcolor}
\usepackage{fancyhdr}

\fancypagestyle{firstpagefooter}{
  \fancyhf{}

  \fancyfoot[C]{\footnotesize Preprint. To appear in the Proceedings of the 2026 IEEE Frontiers in Education Conference (FIE). Copyright 2026 by the author(s).}
}

\def\BibTeX{{\rm B\kern-.05em{\sc i\kern-.025em b}\kern-.08em
    T\kern-.1667em\lower.7ex\hbox{E}\kern-.125emX}}

\begin{document}

\title{Engineering Students' Self-Efficacy, Perceptions, and Performance in a Flipped CS1 Course}

\author{\IEEEauthorblockN{Griffin Pitts}
\IEEEauthorblockA{\textit{North Carolina State University} \\
Raleigh, NC, USA \\
wgpitts@ncsu.edu}
\and
\IEEEauthorblockN{Ashish Aggarwal}
\IEEEauthorblockA{\textit{University of Florida} \\
Gainesville, FL, USA \\
ashishjuit@ufl.edu}}

\maketitle
\thispagestyle{firstpagefooter}

\begin{abstract}
This full research paper investigates how engineering students' course-related beliefs relate to exam performance in a flipped introductory programming course. Understanding factors that influence student learning and performance has long been a focus of computing education research. While prior studies have identified psychological and contextually relevant predictors of success, much of this work has examined students majoring in computer science. Yet introductory programming courses now serve many students from other disciplines, whose beliefs and motivations may differ. To examine these relationships in an engineering-focused CS1 context, we analyze survey and exam data from 602 students. An exploratory factor analysis identified three latent factors: self-efficacy, attitudes toward learning, and perceived programming difficulty. Self-efficacy was positively associated with exam performance, while perceived difficulty was negatively associated. Differences in reported beliefs were also observed across demographic groups, even when performance outcomes were similar. These findings align with and extend prior research, highlighting the role of self-efficacy in achievement and persistence in computing education.
\end{abstract}


\section{Introduction}

Learning a new concept is a complex process involving multiple interrelated factors that can influence one's learning outcome. In computing education, researchers have long examined predictors of learning, performance, and persistence, yet much of this work has been situated within introductory courses designed primarily for students pursuing computer science as a primary field of study. As computing education has expanded and enrollments have diversified to include large numbers of students from other disciplines, evidence suggests that these learners approach programming differently. For example, Dawson et al. \cite{dawson2018designing} reported that non-CS majors in a traditional CS1 course had lower pass rates, weaker expert-like attitudes, and less satisfaction than CS majors, motivating a redesigned course tailored to non-majors. Studies of CS0 and CS1 for non-majors similarly document higher fear and lower confidence that can be mitigated by contextualized, scaffolded instruction (e.g., \cite{ali2025even}). Further, Sax et al. \cite{sax2017examining} found that as non-CS majors increasingly enroll in introductory computing courses, they bring more diverse backgrounds, less prior programming experience, and different motivations for taking these courses compared to CS majors. These differences motivate closer examination of belief constructs among students such as those in engineering who encounter computing in support of their primary field of study.
 
Non-cognitive and psychological factors have been shown to be particularly important in explaining student success in STEM contexts. Self-efficacy, or confidence in one's ability to succeed at a specific task, has been consistently linked to persistence and achievement in both engineering and computer science \cite{hutchison2006factors, loo2013sources, marra2009women}. Research in educational psychology has identified related constructs such as growth mindset \cite{dweck1988social, dweck2000self} and grit \cite{christopoulou2018role}. Much of the intervention work in computing education research has emphasized pedagogical approaches such as flipped classrooms, pair programming, and peer instruction \cite{pair_programming, pair_programming_2, latulipe2018longitudinal, peer_instruction}. However, they represent only one piece of the larger puzzle of learning, as there is a lack of interventions that address these beliefs at a deeper level in the learning process \cite{Malmi_theory}. To understand the complexity of learning, it is essential to recognize that it involves multiple latent constructs that can interact in known and unknown ways. These constructs can include the learner's motivation, prior knowledge, cognitive abilities, and cultural background, among others \cite{lishinski201928}.
 
Therefore, beliefs play an important role in shaping how students engage with course material and persist through educational challenges. Previous work in CS education has examined student beliefs about programming and developed instruments to capture attitudes and perceptions \cite{lishinski2017students}. Given the increasing emphasis on programming across disciplines, it is important to understand how these beliefs influence learning experiences in contexts where programming serves as a supporting skill, such as engineering.

In this paper, we examine the course-related perceptions of engineering students enrolled in an introductory programming course (CS1) and investigate how these perceptions relate to course performance. Using exploratory factor analysis, we identify latent factors representing self-efficacy, attitudes toward learning, and perceived programming difficulty, and test their associations with exam performance. We further analyze how these beliefs vary by gender, prior programming experience, and prior academic achievement. These findings contribute to a deeper understanding of how students' confidence and perceptions influence engagement and achievement in CS1, informing instructional practices that foster self-efficacy and reduce perceived difficulty among engineering students in computing education.
 
\section{Background}
A wide body of research has examined the role of non-cognitive and affective factors in student learning. In education more broadly, emotions, motivation, and self-efficacy have been consistently linked to performance and persistence \cite{pekrun2017achievement, marchand2012role}. Attitudes toward learning, including goal orientation, have been found to further influence how students engage with challenging material and the strategies they employ \cite{nelson2015motivational, lishinski2016learning}. Computer science education research has built on this work by investigating constructs such as self-efficacy \cite{steinhorst2020revisiting, kinnunen2011cs, bandura1999self, schunk2009self, ramalingam2004self}, motivation \cite{zhang2015investigating, mccartney2016computing}, sense of belonging \cite{veilleux2013relationship}, and mindset \cite{gorson2019students}, all of which influence students' experiences in introductory programming courses \cite{multon1991relation, kanaparan2019effect}. For example, Lewis et al. \cite{lewis2011deciding} found that students' self-assessments of their programming ability shaped their persistence. In contrast, Gorson et al. \cite{gorson2019students} observed that students used unexpected criteria, such as typing speed or ease of debugging, to evaluate their programming self-efficacy. Recent work has also explored outcome expectancy, or students' expectations of their own performance, as a related motivational belief \cite{pitts2024outcome}. Variations in students' expectations of their own performance have been associated with differences in self-efficacy, attitudes, and perceptions, and have been found to predict achievement in introductory programming courses \cite{pitts2024outcome}.
 
Although these studies highlight the importance of beliefs and motivation, the expansion of CS0 and CS1 offerings has led to students entering introductory programming with a wider range of backgrounds, experiences, and disciplinary interests \cite{guzdial2017generation}. As learners become more diverse in their preparation and experiences with computing, the influence of non-cognitive constructs may vary depending on students' prior exposure and disciplinary context. For many engineering students, programming serves as a complementary skill rather than a primary area of study, which may affect both their motivation and their perceptions of programming. Examining course-related beliefs within this population is particularly timely given the broader need to integrate computational skills across engineering and interdisciplinary curricula \cite{chem_programming}.

In the context of engineering education, Usher et al. found that \textit{``although students may find engineering challenging, their perceived efficacy might benefit from the positive emotions they feel when engaged in their work"} \cite{asee_peer_24723}. In 2020, Malmi et al. conducted a comprehensive survey of the theories and affective constructs used in the domain of CS education \cite{Malmi_theory}. They proposed \textit{``that an improved understanding of the complex factors influencing learning related to students' internal factors such as beliefs, emotions, and attitudes, would help us to develop more effective interventions to influence students' perceptions about learning programming, and thus to improve pass rates and learning results."} \cite[p2]{Malmi_theory}. 

Prior analysis of engineering students in this flipped CS1 context examined the influence of gender, prior academic achievement, and prior programming experience on students' learning self-efficacy and outcome expectancy \cite{aggarwal2023identifying}. The study found that learning self-efficacy and prior academic achievement were significant predictors of students' expected final grades. Building on this, this paper examines the course-related beliefs of engineering students in an introductory programming course, with a focus on group-based differences and their relationship to performance. Understanding how these beliefs vary across demographic and experiential factors can provide insight into which student groups may benefit from additional support in introductory programming courses.

\subsection{Mindset Theory}
Within the broader set of belief constructs that influence student learning, one that has received sustained attention is the concept of mindset. A growth mindset reflects the belief that abilities can be improved through effort and practice, while a fixed mindset assumes that the ability is largely innate and unchangeable \cite{dweck1988social,dweck2000self,dweck2006mindset}. Research in educational psychology has shown that students with a growth mindset demonstrate higher motivation and stronger academic performance because they are more likely to persist through challenges and setbacks \cite{blackwell2007implicit, brougham2017impact, o2014brain}. Moreover, interventions can influence students' mindsets, suggesting that these beliefs are not static but open to change \cite{blackwell2007implicit, yeager2016using}.
 
However, in computing education, the work on mindset remains relatively limited, and the results are mixed. Simon et al. \cite{simon2008saying} and Cutts et al. \cite{cutts2010manipulating} found that interventions could shift students' mindset beliefs, but these changes did not necessarily lead to measurable improvements in performance. More recently, Gorson et al. \cite{gorson2019students} highlighted that students' self-described mindsets often do not align with the theoretical definitions in the literature, raising questions about how mindset operates in CS1 contexts. Overall, these findings suggest that mindset holds promise for understanding student beliefs in computing; however, our understanding of its development and influences in these courses is still emerging. This perspective motivates the focus of the present study on identifying the belief constructs of engineering students in CS1 through exploratory factor analysis, allowing us to examine how such constructs are organized in practice and how they relate to students' performance.

\section{Course Context}

The data used in this analysis comes from an introductory programming course for engineering students taught at a research-intensive university in the southeastern United States. Over four semesters, the course consisted of multiple sections, all of which followed a flipped classroom model. Multiple sections of the course were offered in both in-person and online formats. Each section had about 50 students. With the flipped classroom model of the course, students were expected to both watch prerecorded module lectures and complete a weekly graded quiz before class. In-person class time was primarily reserved for reviewing the content taught in the video lectures and completing three programming problems with the instructor and peers. Students were expected to work on module homework assignments outside of class, which were typically due at the end of each week. The programming concepts covered in the course included input and output, program control flow, vectors, strings, images, and functions taught in MATLAB. The course content was divided into 14 modules with eight programming-related homework assignments and two exams (a midterm and a final). 

\section{Methodology}

This study was designed to examine the relationship between engineering students' course-related beliefs and their performance in a CS1 course, as well as differences in these beliefs across student groups. Specifically, we addressed the following research questions:

\begin{itemize}
    \item \textbf{RQ1: How are engineering students' course-related beliefs related to their performance in a CS1 course?}
    
    \item \textbf{RQ2: How do these course-related beliefs differ across students' gender, prior programming experience, and prior academic achievement?}
\end{itemize}

To address these questions, data was collected across four semesters from students enrolled in a CS1 course. During the first week of the semester, students completed a survey capturing demographic characteristics, prior experiences, and course-related beliefs. Student performance was assessed through the average of two course exams, which provided a standardized measure of achievement. The following sections describe the survey instrument, participant demographics, and analytic procedures in more detail.

\subsection{Survey Instrument}
  
The survey consisted of two sections. The first section collected demographic and academic information, including gender, prior programming experience (PPE), academic year, prior academic achievement, intended field of study, as well as course-related information such as whether the student was concurrently enrolled in the associated lab and their expected final grade. Students' prior programming experience (PPE) was categorized as either 0 (no substantial experience) or 1 (some experience), determined based on their self-reported experience level and written descriptions of past exposure to programming. Students who reported no prior programming experience were assigned a 0. For those with minimal reported experience, written explanations were manually reviewed using established criteria: students received a 0 if they had only block-based coding experience, described their knowledge as limited or basic, or used programming minimally in a non-programming context. Conversely, students received a 1 if they mentioned specific programming languages, had completed or partially completed a formal programming course, or programming-related certifications. Ambiguous cases were resolved at the researchers' discretion.

The second section of the survey measured students' beliefs related to programming and learning in the course. Twelve Likert-scale items (5-point scale from “Strongly Disagree” to “Strongly Agree”) were designed to capture students' level of agreement on the following questions:
\begin{itemize}
     \item \textit{"I am excited and looking forward to taking this course"}, 
     \item \textit{"I am good with computers"}, 
     \item \textit{"I care more about learning than earning good grades"},
     \item \textit{"I feel confident in my ability to learn programming"},
     \item \textit{"I feel confident in my ability to solve problems"},
     \item \textit{"I generally attend classes even if the attendance is not mandatory"},
     \item \textit{"I know that watching module videos before coming to class every week will be critical for my success"},
     \item \textit{"I prefer online classes rather than in-person classes"},
     \item \textit{"I think I will do great in this course"},
     \item \textit{"I think it is difficult to learn programming"},
     \item \textit{"I think learning to program involves a lot of effort"},
     \item \textit{"Programming is a useful skill"}
\end{itemize}

These items were designed to measure students' beliefs about programming, confidence in their ability to succeed, and attitudes toward learning behaviors that could influence performance. While item wording was tailored to the flipped CS1 engineering context, it was informed by established instruments including the Computer Science Attitude Survey \cite{wiebe2003computer}, and the Attitude Scale of Computer Programming Learning \cite{korkmaz2014validity}. Since the survey combined items adapted from multiple prior instruments and was newly tailored to the study's instructional context, exploratory factor analysis was conducted to examine the relationships among items and to identify the latent belief constructs underlying students' responses.

\begin{table*}[h]
    \centering
    \caption{Exploratory Factor Analysis of the Survey Instrument}
    \label{tab:factorLoadings}
    \begin{tabular}{p{0.5\textwidth}cccc}
        \toprule
        Item & Factor 1 & Factor 2 & Factor 3 & Uniqueness \\
        \midrule
        I think I will do great in this course & 0.885 & & & 0.198 \\
        I feel confident in my ability to learn programming & 0.901 & & & 0.202 \\
        I feel confident in my ability to solve problems & 0.714 & & & 0.513 \\
        I am good with computers & 0.500 & & & 0.662 \\
        I am excited and looking forward to taking this course & & 0.565 & & 0.441 \\
        I care more about learning than earning good grades & & 0.463 & & 0.727 \\
        I generally attend classes even if the attendance is not mandatory & & 0.468 & & 0.752 \\
        I know that watching module videos... will be critical for my success & & 0.748 & & 0.436 \\
        I think learning to program involves a lot of effort & & & 0.835 & 0.288 \\
        I think it is difficult to learn programming & & & 0.753 & 0.346 \\
        I prefer online classes rather than in-person classes & & & & 0.931 \\
        Programming is a useful skill & & & & 0.680 \\
        \midrule
        Eigenvalue & 3.856 & 2.264 & 1.216 & \\
        Variance explained (\%) & 21.8 & 13.1 & 10.8 & \\
        Cumulative variance (\%) & 21.8 & 34.9 & 45.7 & \\
        \bottomrule
        \addlinespace[1ex]
        \multicolumn{5}{p{0.55\textwidth}}{\textit{Note.} Applied rotation method is oblimin. Factor loadings $< 0.4$ are suppressed.} \\
    \end{tabular}
\end{table*}

\subsection{Participants} 
  
The participation period spanned four semesters, including 620 students. 18 students were excluded from the data set due to missing information, including late withdrawal from the course or failure to take the final exam, resulting in a sample size of 602 students. Overall, the most common intended fields of study were Mechanical Engineering (169 students; 27.66\%), Aerospace engineering (104 students; 17.02\%), Civil and/or Environmental Engineering (114 students; 18.66\%), and Biomedical Engineering (103 students; 17.11\%), and the remaining students belonged to a variety of other disciplines (108 students; 19.55\%). Of the 602 students, 239 students (39.70\%) had some relevant form of prior programming experience, while the remaining 363 students (60.30\%) had no or minimal experience. There were 382 male students (63.46\%), 214 female students (35.55\%), and six (0.99\%) who preferred not to disclose or self-described. There were 166 first-year students (27.17\%), 376 second-year students (62.55\%), 41 third-year students (6.6\%), and 19 students who were in their fourth academic year or beyond (2.9\%). The average self-reported grade point average was a 3.626 out of 4.0 (SD: 0.343). The exam average was an 81.95. The majority of students (71.26\%) indicated prior experience taking a course using a flipped classroom model.

\subsection{Exploratory Factor Analysis}

To identify the latent belief constructs underlying students' survey responses, we conducted Exploratory Factor Analysis (EFA). EFA identifies shared variance among variables and combines them into factors meant to represent the underlying relationships between measured items.

Before proceeding with EFA, we verified its applicability through both Bartlett's Test of Sphericity and the Kaiser-Meyer-Olkin (KMO) test. These tests assess whether factor analysis is appropriate by examining the significance of correlations within the variable matrix and evaluating the strength of relationships between variables. The Bartlett's test was significant ($\chi^2$ = 2812.083, p $<$ .001) and the KMO value confirmed sufficient sampling adequacy, indicating that the data were suitable for exploratory factor analysis and that the items shared enough common variance to justify the extraction of latent factors. In conducting the EFA, we employed principal axis factoring with oblimin rotation, as we expected the underlying factors to be correlated. Items were assigned to a factor if their loading was $\geq 0.40$. The number of factors was determined using eigenvalues greater than 1, inspection of the scree plot \cite{cattell1966scree}, and theoretical interpretability. Three factors were retained, which explained 45.7\% of the variance.

\subsection{Further Analysis}
  
To examine the relationship between belief factors and performance, Pearson's correlation coefficients were calculated between factor scores and exam averages. To further explore these relationships, students were divided into quartiles based on exam scores (Q1 = lowest performers, Q4 = highest performers), and Kruskal–Wallis tests were used to compare factor scores across quartiles. This approach allowed us to identify trends that might not be evident in correlations alone. 

To analyze group differences, nonparametric tests were used due to non-normal distributions of some factor scores. Mann–Whitney U tests compared factor scores by gender and prior programming experience. Pearson's correlations were used to examine the relationships between prior academic achievement of the students (GPA) and their course-related beliefs, as well as between GPA and exam performance.

\section{Results}

\subsection{Factors Derived from Exploratory Factor Analysis (EFA)}
  
Through EFA, as described in the methodology section, three distinct factors were identified that optimally describe the constructs present in the survey data. A chi-squared test for model fit was significant (${\chi}^2$ = 260.419, df = 42, p $<$ .001), indicating that our factor structure adequately represents the underlying patterns in the data (see Table \ref{tab:factorLoadings}).

The three-factor solution explained 45.7\% of the total variance in the data. The first factor, labeled \textbf{self-efficacy}, accounted for 21.8\% of the variance and included items such as ``\textit{I feel confident in my ability to learn programming}'' (0.901) and ``\textit{I think I will do great in this course}'' (0.885). The second factor, \textbf{attitude toward learning}, explained 13.1\% of the variance and consisted of items reflecting students' general learning approaches and course-specific attitudes. The third factor, \textbf{perceptions about programming}, explained 10.8\% of the variance and included items relating to programming difficulty, such as ``\textit{Learning to program requires a lot of effort}'' (0.835) and ``\textit{Programming is difficult to learn}'' (0.753). Two items did not meet the loading threshold and were excluded from the final factor structure.

\subsection{RQ1: Association between Factors and Performance}

To address RQ1 (How are engineering students' course-related beliefs related to their CS1 performance?), we first conducted correlation analyses to determine the direct relationships between each factor and performance, followed by an analysis of how these factors were distributed and varied between different performance quartiles. 

\subsubsection{Correlations:} Pearson's correlations showed that self-efficacy was positively related to exam performance (r = 0.081, p = .044), while perceptions of programming difficulty were negatively related to performance (r = -0.079, p = .050). The attitude toward learning factor showed a small positive trend with performance, but did not reach statistical significance (r = 0.035, p = 0.392).

\subsubsection{Quartile analyses:} To further investigate, students were divided into quartiles based on exam scores (Q1 = lowest performers, Q4 = highest performers). This quartile-based approach allowed for comparisons across performance groups, revealing non-linear trends or threshold effects that might not be apparent through simple correlations. Kruskal–Wallis tests indicated significant differences across quartiles for self-efficacy ($\chi^2$(3) = 10.662, p = .014) and perceptions of programming difficulty ($\chi^2$(3) = 12.177, p = .007), but not for attitude toward learning ($\chi^2$(3) = 3.849, p = .278). 

Self-efficacy scores increased across performance quartiles, with Q4 students (highest performers) reporting the strongest self-efficacy (M = 4.150, SD = 0.702) compared to Q1 students (lowest performers; M = 3.908, SD = 0.768). In contrast, perceptions of programming difficulty were higher among lower-performing quartiles. Q1 (M = 4.090, SD = 0.827) and Q2 (M = 4.109, SD = 0.921) reported the greatest perceived difficulty, Q3 reported the lowest (M = 3.830, SD = 0.981), and Q4 showed intermediate values (M = 3.946, SD = 0.851).

\subsection{RQ2: Differences within Observed Belief Factors Based on Student Characteristics}

\subsubsection{Gender}
  
Analysis of gender differences revealed significant variations in belief factors. Male students reported significantly higher levels of self-efficacy (M = 4.193, SD = 0.612) than female students (M = 3.877, SD = 0.646), with this difference being statistically significant (F(1,594) = 35.057, p $<$ .001). Conversely, female students reported higher perceptions of programming difficulty (M = 4.201, SD = 0.610) compared to male students (M = 3.978, SD = 0.823), also statistically significant (F(1,594) = 12.028, p $<$ .001). No significant gender differences were found in attitudes toward learning (F(1,594) = 2.474, p = 0.116). Despite these differences in beliefs, there were no significant differences in exam performance (t = -0.842, p = 0.400).

\subsubsection{Prior Programming Experience}
  
Students with prior programming experience (PPE) demonstrated significantly higher self-efficacy (M = 4.274, SD = 0.570) compared to those without experience (M = 3.950, SD = 0.656), with F(1,600) = 39.006, p $<$ .001. Students without prior experience reported higher perceptions of programming difficulty (M = 4.205, SD = 0.634) than those with experience (M = 3.822, SD = 0.881), a difference that was statistically significant (F(1,600) = 38.430, p $<$ .001). Both groups showed similar attitudes towards learning, without significant differences (F (1,600) = 0.355, p = 0.552). Unlike with gender, there was a significant difference in exam performance between students with and without PPE (t = -2.827, p = 0.005).

\subsubsection{Grade Point Average}
Our analysis did not find significant correlations between prior academic achievement of the students (GPA) and two of the course-related beliefs: self-efficacy (r = 0.021, p = 0.615) and perceptions about programming (r = 0.004, p = 0.920). A significant correlation was observed between GPA and attitudes toward learning (r = 0.099, p = 0.016), and there was a significant correlation between GPA and exam average (r = 0.512, p $<$ 0.001).

\section{Discussion}
This study examined how engineering students' course-related beliefs (i.e. self-efficacy, attitudes toward learning, and perceptions of programming difficulty) relate to their performance in a flipped CS1 course. The findings extend prior computing education research, and focus on engineering students, who enter programming courses with distinct motivations, experiences, and attitudes toward learning to program.

\subsection{Self-Efficacy and Performance}

Consistent with social cognitive theory \cite{bandura1999self} and prior CS education research, self-efficacy emerged as a meaningful predictor of engineering student outcomes. Students with higher self-efficacy reported stronger performance, echoing findings that self-efficacy beliefs shape persistence and engagement in computing \cite{ramalingam2004self, kinnunen2011cs, multon1991relation}. Although the correlation between self-efficacy and exam performance was modest, the quartile analyses revealed a clear pattern, where high-performing students reported stronger confidence in their programming abilities. This supports earlier work showing that students' self-assessments of their programming ability can strongly influence whether they persist in CS1 \cite{lewis2011deciding}.

However, gender-based differences in self-efficacy persist despite equivalent performance, with female students reporting lower confidence than their male peers. This finding aligns with multi-institutional studies of engineering and computing students, which have consistently documented gender disparities in self-efficacy \cite{marra2009women, hutchison2006factors}. Importantly, these differences in beliefs, rather than ability, may influence female students' experiences and sense of belonging in programming \cite{veilleux2013relationship}.

\subsection{Perceptions of Programming Difficulty}

Students' perceptions of programming difficulty were negatively associated with exam outcomes. Those who viewed programming as especially difficult were more likely to fall into lower performance quartiles, reflecting prior work linking difficulty perceptions to disengagement and attrition in introductory computing courses \cite{peteranetz2018helping, pirttinen2020study}. Prior experiences and motivations influence how students approach and perform in introductory programming courses. Indeed, students with prior programming experience reported both higher self-efficacy and lower perceptions of difficulty, consistent with earlier studies demonstrating the powerful influence of prior exposure on beliefs and outcomes \cite{guzdial2017generation, gorson2020cs1}.

Mindset theory posits that students who believe that their abilities can be developed through effort (a growth mindset) are more likely to persist through challenges than those who view ability as fixed \cite{dweck2000self, dweck2006mindset}. The existence of a positive association between self-efficacy and performance observed in this study is aligned with the mindset theory framework, suggesting that the confidence of students in their ability to learn programming can reflect or reinforce growth-oriented beliefs. In contrast, higher perceptions of programming difficulty among students with lower performance may signal more fixed views of ability, where challenges are seen as limitations rather than opportunities for improvement. Although mindset was not directly measured, the relationship between self-efficacy, perceived difficulty, and performance highlights the broader influence of students' belief systems on motivation and engagement in introductory computing courses. Future work could explicitly examine how mindset and self-efficacy interact to influence persistence in programming contexts.

\subsection{Attitudes Toward Learning and Course Engagement}

Attitudes toward learning showed weaker associations with performance in this study, though the positive trend is consistent with earlier work linking motivation and goal orientation to programming outcomes \cite{lishinski2016learning, nelson2015motivational}. One possible explanation is that while broad learning orientations are important, more proximal behaviors, such as engaging with flipped classroom materials, may mediate the relationship between attitudes and outcomes. Recent work on flipped CS1 classrooms found that students' engagement with pre-class lectures, perceived availability of support, and quality of in-class activities were strongly related to exam performance, even after accounting for self-efficacy and demographics \cite{pitts2025flipped}. While general learning attitudes contribute to performance, the day-to-day behaviors and perceptions students bring into a flipped classroom may be equally or more influential. 

\subsection{Implications for Pedagogy and Intervention}
These results point toward several avenues for instructional support. First, interventions that target confidence-building may help address persistent self-efficacy gaps across gender and experience groups. Techniques such as peer instruction \cite{peer_instruction}, pair programming \cite{pair_programming, pair_programming_2}, and flipped classrooms \cite{latulipe2018longitudinal} have been shown to support self-efficacy and engagement, and may be particularly beneficial for students who perceive programming as intimidating. Second, designing early course activities that provide mastery experiences could reduce miscalibrated perceptions of difficulty, particularly among students without prior programming experience \cite{loo2013sources, asee_peer_24723}. Finally, given that attitudes toward learning are only weakly predictive on their own, instructors may need to scaffold engagement in flipped contexts, ensuring students have structured opportunities to build confidence and programming experience.  

\section{Limitations \& Future Work} 

This study was conducted in the context of a flipped CS1 course for engineering students, and the findings should be interpreted with that setting in mind. Belief patterns and their relationship to performance may look different in other types of courses or among students from different majors. Another limitation concerns the way prior programming experience was measured. For our analysis, experience was coded as a binary variable, but in reality programming background is not discrete and can vary in both depth and context. While this allowed us to examine broad differences, it does not capture the full range of student preparation.

In addition, our survey instrument relied on context-specific items rather than established scales for constructs such as self-efficacy or attitudes. This limited direct comparison with other studies but motivated the use of exploratory factor analysis to identify the latent belief constructs present in the data. Future work should consider more detailed measures of programming experience and incorporate a mix of validated scales and context-specific items to allow both comparability across studies and sensitivity to the unique context of engineering students in CS1. Extending this analysis to other institutional settings and student populations would further clarify how these course-related beliefs vary and how they can inform the design of more inclusive and supportive introductory programming courses.

\section{Conclusion}
  
This study examined the course-related beliefs of engineering students enrolled in a flipped CS1 course. Using exploratory factor analysis of survey data, we identified three latent constructs (self-efficacy, attitudes toward learning, and perceptions of programming difficulty) that captured the underlying dimensions of students' beliefs about learning to program. The findings indicate that although these beliefs are linked to performance, they also vary by gender and prior programming experience, revealing disparities that are not captured by achievement alone. These patterns highlight the importance of examining technical outcomes alongside how students perceive and experience programming. For educators and researchers, creating inclusive introductory courses requires refining pedagogy, attending to the beliefs students bring with them, fostering confidence, and providing opportunities that account for varied backgrounds and experiences. By examining engineering students' non-cognitive and psychological factors in computing education, this study highlights the role of confidence and perceived difficulty in persistence. Considering these constructs alongside classroom engagement and learning behaviors can inform interventions that support success in introductory programming.

\bibliographystyle{ieeetr}
\bibliography{sample-base}

\end{document}